\documentclass[12pt]{article}  

\usepackage{amsmath,amsfonts,amssymb}
\usepackage{graphicx}
\usepackage{float}
\usepackage[colorlinks=true, allcolors=blue]{hyperref}

\providecommand{\keywords}[1]
{
  \small	
  \textbf{\textit{Keywords---}} #1
}

\title{A pyramid-based adaptive optics for the high-resolution echelle spectrograph at SAO RAS 6-m telescope}

\author{
{Muslimov E.}$^{a,b,c}$,{Valyavin G.}$^{d}$,{Chambouleyron V.}$^{b}$ ,\\
{Pedreros F.}$^{b}$, {Boudjema I.}$^{b}$\\
\small a--{NOVA-ASTRON Oude Hoogeveensedijk 4,}\\
\small {7991 PD Dwingeloo, The Netherlands}\\
\small b--{Laboratoire d'Astrophysique de Marseille, France}\\
\small c--{Kazan National Research Technical University}\\
\small {named after A.N. Tupolev KAI, 10 K. Marx, Kazan, Russia, 420111}\\
\small d --{Special Astrophysical Observatory of Russian Academy of Science,}\\
\small {Nizhnij Arkhyz, Russia 369167}\\
}

\begin{document} 
\maketitle

\begin{abstract}
We propose a design of an adaptive optics (AO) system for the high-resolution fiber-fed echelle spectrograph installed at the Nasmyth focus of the 6-m BTA telescope at the Special Astrophysical Observatory (SAO) of the Russian Academy of Sciences (RAS). The system will be based on a pyramid wavefront sensor and benefit from the experience of the Laboratoire d’Astrophysique de Marseille team in the field of adaptive optics.  The AO will operate in the visible domain of 430-680 nm, in an f/30 input beam and provide correction for the on-axis source only. The main challenges in this particular design are insetting  inserting the AO into an existing optical system and maintaining the focal and pupil planes configuration, fitting within the instrument’s flux budget as well as limitations on the total cost of the AO bench. According to the current design, the AO bench will use an additional relay consisting of 2 spherical mirrors to re-collimate the beam and project the pupil onto a small deformable mirror. A dichroic splitter will be used to longwave component to the pyramid wavefront sensor branch based on refractive optics only. Using off-the-shelf components only we can reach the instrumental wavefront error of 0.016 waves PTV with a 20 nm bandpass filter at 700 nm. Using folding mirrors and refocusing of the fiber’s microlens we restore the nominal geometry of the beam feeding the spectrograph. The final goal for the AO system is to increase the energy concentration in spot at the spectrograph’s entrance, and our preliminary modelling shows that we can gain by factor of 69.5 with the typical atmospheric conditions at SAO RAS.
\end{abstract}


\section{INTRODUCTION}
\label{sec:intro}  

The pyramid wavefront sensor was proposed
for the first time in Ragazzoni \cite{Ragazzoni96}, is an optical device used to perform wavefront sensing by
performing optical Fourier filtering with a glass pyramid
with four sides that is located at the focal plane. The purpose
of this glass pyramid is to split the incoming beam into four beams producing four different filtered images
of the entrance pupil. This filtering operation allows the conversion of phase information at the entrance pupil into amplitude at the pupil plane, where a quadratic sensor is used to record the signal \cite{Verinaud04,Guyon05}. Today the pyramid wavefront sensors are on a high demand in the astronomical applications. The main advantage of them is a high sensitivity, superior over that of the Shack-
Hartmann wave-front sensor (WFS) \cite{Esposito01}, while the main downside is their non-linearities, which that prevent a
simple relation between the incoming phase and the measurements, leading to control issues in the adaptive optics (AO) loop. However, the latter issue could be solved by using the optical gain tracking technique \cite{Chambouleyron21}.

Recently it was successfully demonstrated that a fully-functional AO bench based on a pyramid WFS can be developed and commissioned with moderate resources. The PAPYRUS project \cite{Muslimov21} was designed by a team of young researchers at Laboratoire d'Astrophysique de Marseille and installed at 1.52 meter telescope at Observatoire de Haute Provence. The bench uses only existing and off-the-shelf components was created in less than two years within a limited budget. Both in-lab and the first on-sky performance metrics are in a good agreement with the modelling predictions.  

The success of PAPYRUS has urged us to look for potential applications, where a similar AO system could be introduced within a moderate budget and in a relatively short time, but with a notable scientific outcome. One of the promising options consists of development of a similar system to feed an echelle spectrograph at 6-m telescope. The spectrograph works with a fiber input, so the AO loop should correct only the on-axis point. Also, the spectral working range of the instrument is not so wide, so the requirements for the chromatic aberration correction in the AO system can be relaxed. In the meantime, using the AO could significantly increase the energy concentration at the spectrograph's input, thus increasing its' sensitivity. 

So in the present paper we consider optical design of a pyramid-based AO system for a high-resolution echelle spectrograph at 6-m telescope. Below we introduce the target spectral instrument, then we discuss the adaptive optics system design and present the expected performance in the WFS and science branches and its' impact on the instrument capacity.

\section{ECHELLE SPECTROGRAPH}
\label{sec:echelle}  

The target instrument is the fiber-fed high-resolution echelle spectrograph, recently built for at Special Astrophysical Observatory of Russian Academy of Science (SAO RAS) \cite{Valyavin14}. The instrument is designed for the 6-m alt/az mounted telescope (BTA - Big Telescope Alt-azimuth) and has the following key features:
\begin{itemize}
    \item Spectral resolving power up to R = 100 000 with a possibility (to use lower resolution modes by using pixel binning;
    \item The simultaneously detected waveband of 400-750 nm;
    \item Auxiliary units to measure the Stocks parameters, perform photometric and spectral calibrations;
    \item The light from the telescope focus is fed to the spectrograph through an optical fiber.
\end{itemize}

The spectrograph is intended for Doppler studies of exoplanets and multiple stellar systems. Its' science goals also include studies of stellar atmospheres, asteroseismological studies, studies of stellar magnetism, active nuclei of bright galaxies and interstellar medium.

\begin{figure} [ht]
   \begin{center}
   \begin{tabular}{c} 
   \includegraphics[width=12cm]{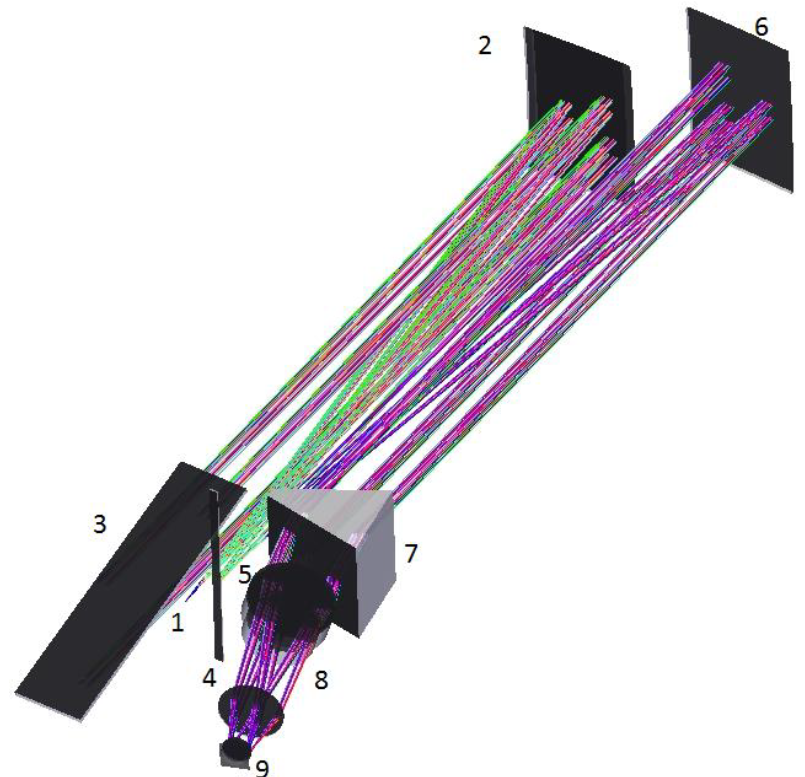}
   \end{tabular}
   \end{center}
   \caption[echelle] 
   { \label{fig:echelle} 
Optical design of the echelle spectrograph: 1 - entrance slit, 2, 6 - collimating mirrors, 4 - folding mirror, 5 - field stop, 7 - cross-disperser, 8 - camera lens, 9 - CCD detector}
   \end{figure}

The spectrograph optical design is shown in Fig.~\ref{fig:echelle}. The light emitted from the entrance slit 1 is collimated by an F/11.6 mirror 2 and incident onto the echelle grating 3, which operates close to the auto-collimation mode together with the collimator. After the second reflection from 2 the dispersed beam is folded by the flat mirror 4 and is collimated again by the transfer collimator 6. It its' focal plane 5 the white pupil is formed, and in this plane the cross-disperser grism 7 is mounted. Further, the beam is focused by the camera lens 8 onto the CCD detector 9.
  
The key inputs for the AO system design are:
\begin{itemize}
    \item Installation at the Nasmyth focus with f/30;
    \item Correction only for the on-axis point;
    \item The working spectral range is at least 430-680 nm (400-750 nm goal);
    \item The allowable fraction of flux used in the AO branch - 10\%;
    \item The entire AO system should pick-up the beam and then return it to the nominal path with the same f/\# and the focus position;
    \item The distance between AO pick-off mirror and the focal plane is 400 mm.
    
On top of this it is preferable to use off-the-shelf components and rely on reflective optics.    

\end{itemize}

\section{ADAPTIVE OPTICS SYSTEM DESIGN}
\label{sec:AO}  

In the optical design of the AO bench we tried to re-use the PAPYRUS heritage and rely on the same or similar active components, namely:
\begin{enumerate}
    \item Deformable mirror (DM) with at least 17x17 actuators and 37.5 mm clear aperture diameter;
    \item WFS branch camera with $5.76 \times 5.7 mm^2$ sensing area of $240\times240$ pixels;
    \item The glass pyramid with the following parameters: facet angle $8.9^{\circ}$, material LF5, leading to deflection angle of $\pm 5.44^{\circ}$;
    \item Off-the-shelf modulation mirror with 12 mm diameter;
\end{enumerate}

\begin{figure} [ht]
   \begin{center}
   \begin{tabular}{c} 
   \includegraphics[width=13cm]{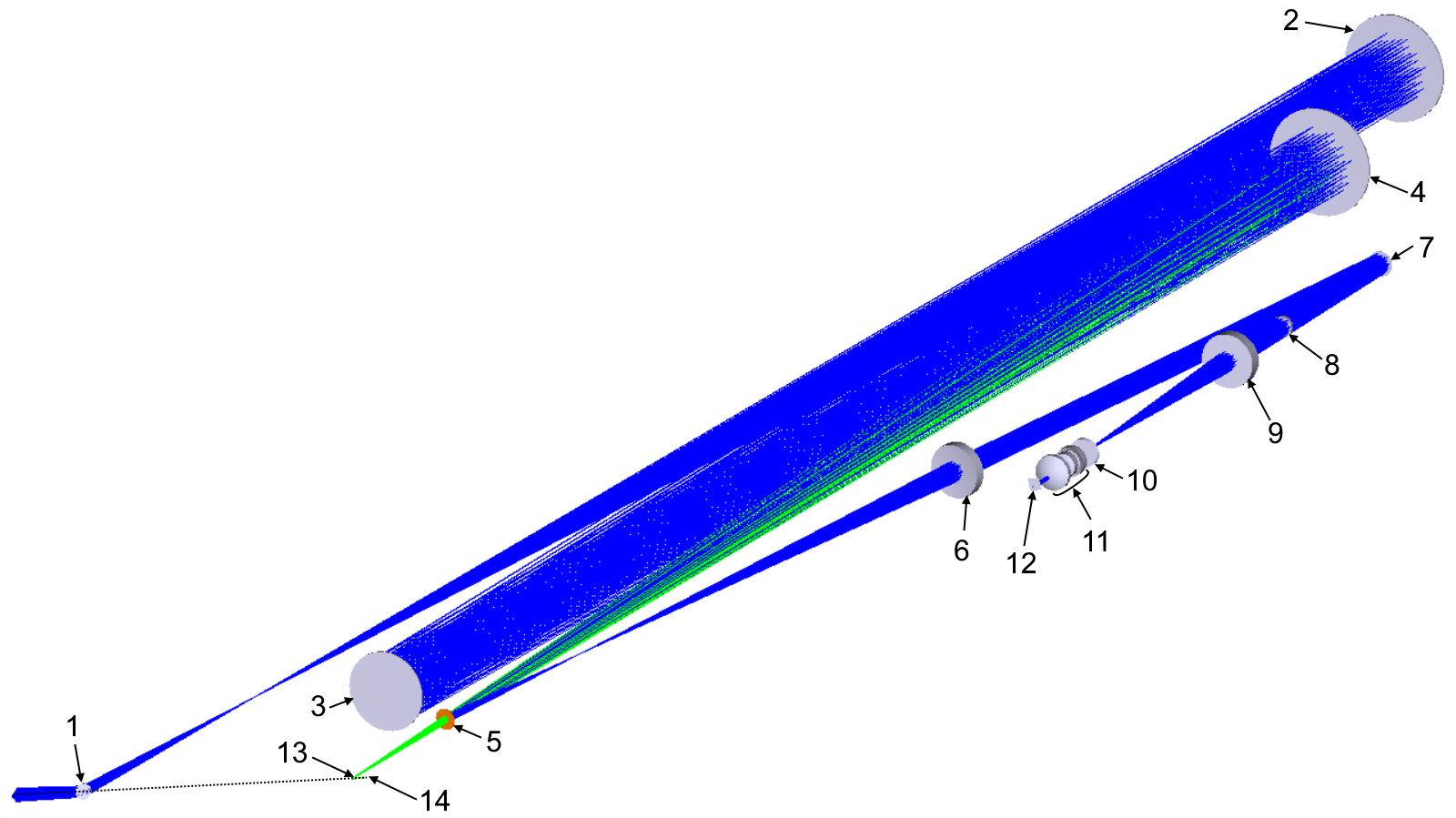}
   \end{tabular}
   \end{center}
   \caption[ao] 
   { \label{fig:ao} 
Optical design of pyramid adaptive optics system: 1 - pick-off mirror, 2, 4 - spherical mirrors relay, 5 - deformable mirror, 5 - dichroic splitter, 6, 9 - wavefront sensing branch lens relay, 7 - modulation mirror, 8 - filter, 10 - pyramid, 11 - WFS camera lens, 12 - WFS CCD, 13 - folding mirror, 14 - spectrograph fiber end.}
   \end{figure}

The proposed optical design is shown in Fig.~\ref{fig:ao}. The incoming beam is sent to the AO system by the folding mirror 1 and collimated by mirror 2. Since the beam is relatively slow it is sufficient to use commercial 2-inches spherical mirror with f=1000 mm. The pupil is reconstructed on th DM 3. The reflected beam is focused again by the mirror 4 identical to 2. Note that the mirrors 2 and 4 are slightly shifted to provide the necessary pupil projection and  facilitate their mounting. The science and the WFS branches are separated by the dichroic splitter 5, which allows to minimize the flux losses for the scientific payload. It reflects the longer wavelengths right outside of the spectrograph's working range $\lambda>680nm$ and steers it to the collimating doublet lens 6 (d=1 inch, f=300 mm), which reconstructs the pupil again on the modulating mirror 7. Further we use bandpass filter 8 ($\lambda=693-712 nm$)mounted in a collimated beam, to moderate the chromatic aberrations of the WFS branch. The beam is focused by a similar doublet lens 9 to the intermediate focal plane, where the glass pyramid 10 is installed. The 4 beams created by the pyramid are collimated by a commercial short focal length lens 11 ($Canon^{TM}$ EF-S 24mm F/2.8 STM pancake photographic lens\cite{CANON}) and the pupil images are detected by the CCD 12. The shorter wavelengths are transmitted by the dichroic splitter to the spectrograph entrance, and the beam is folded again by mirror 13 to restore the nominal position of the focal plane 14.  

\section{PERFORMANCE ANALYSIS}
\label{sec:performance}  

Below we analyze the static instrumental aberrations, introduced by this simple optical design separately in the science and wavefront-sensing branches. We assume that since we have the DM in the optical train, it is possible to use a part of its' stroke to compensate these static aberrations, as it was successfully demonstrated during the PAPYRUS integration. Finally, we provide a coarse estimate of the AO system performance with the typical atmospheric conditions for the SAO RAS site.

\subsection{Pyramid AO branch}
\label{sec:pyr}

\begin{figure} [ht]
   \begin{center}
   \begin{tabular}{c} 
   \includegraphics[width=9cm]{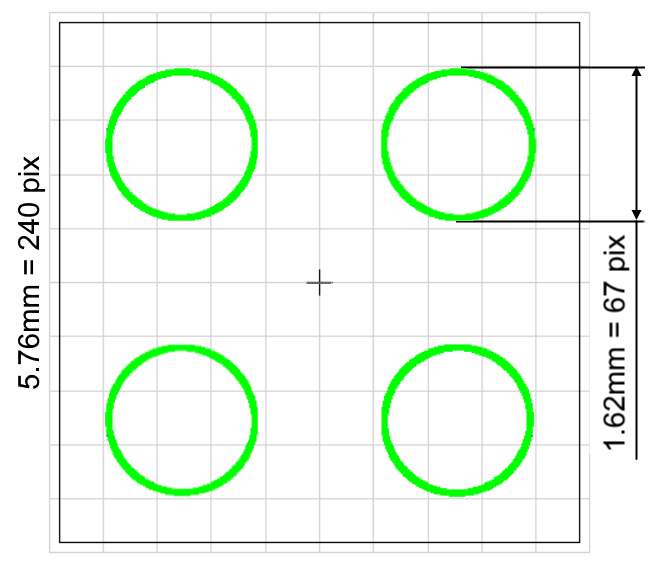}
   \end{tabular}
   \end{center}
   \caption[fillWFE] 
   { \label{fig:fillWFE} 
WFS camera filling and pupil image sampling.}
   \end{figure}

The pyramid creates 4 images of the pupil, which are focused on the WFS CCD (see Fig.~\ref{fig:fillWFE}). Each of the pupil images covers 67 pixels in diameter, which give us to 3.9 or 2.8 pixels per actuators for the 17x17 and 24x24 actuators patterns, respectively. These actuators numbers correspond, for instance, to the commercial DM models DM241 and DM468 by $ALPAO^{TM}$\cite{ALPAO}. They have close clear apertures of 37.5 and 33 mm, respectively, and in both of the cases the projection optics provides a sufficient sampling.

\begin{figure} [ht]
   \begin{center}
   \begin{tabular}{c} 
   \includegraphics[width=13cm]{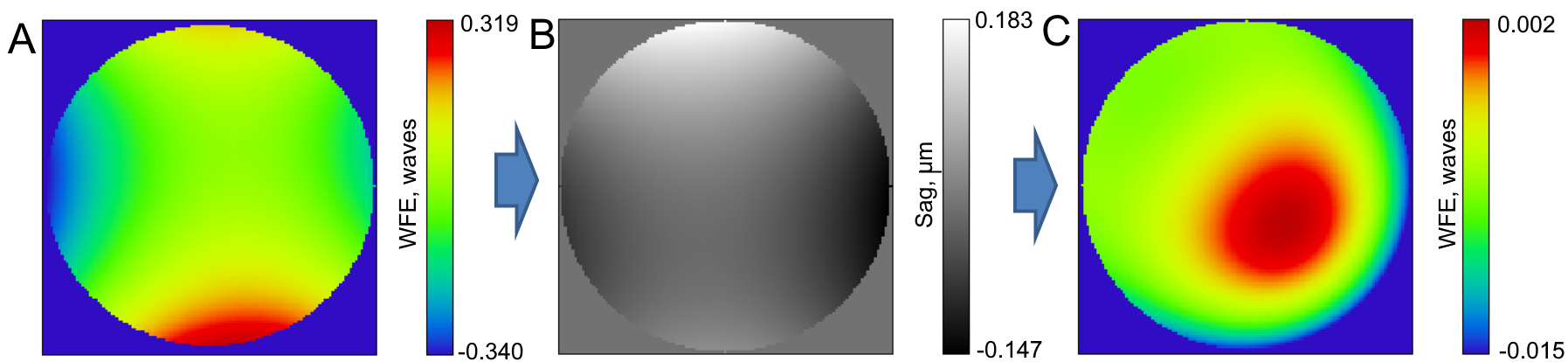}
   \end{tabular}
   \end{center}
   \caption[instrWFE] 
   { \label{fig:instrWFE} 
WFE in the pyramid branch: A - initial WFE with the projection systems residual aberrations, B - DM sag shape to correct the static aberrations, C - corrected WFE in the pyramid.}
   \end{figure}

The simplifications in the optical design as use of the spherical mirrors cause notable residual aberrations. The residual wavefront error (WFE) at 700 nm is shown in Fig.~\ref{fig:instrWFE},A. However, this WFE defined mainly by the spherical aberration and primary astigmatism can  be relatively easily compensated by the DM shape. By optimizing the WFE in the pyramid branch we determine the DM shape necessary for the static aberrations compensation (see Fig.~\ref{fig:instrWFE},B).The peak-to-valley stroke used in this case is only $0.33 \mu m$, while the full stroke for the commercial DM's can reach $12$ or $40 \mu m$ depending on the model \cite{ALPAO}. By applying this small correction we can decrease the instrumental from $0.659\lambda$ to just $0.016 \lambda$ PTV as it is shown in Fig.~\ref{fig:instrWFE},C.

We assume that protected silver coating is used on all of the auxiliary mirrors, and a standard protected aluminium is imposed on the DM, use a typical shortpass dichroic splitter transmission/reflection curve \cite{Thorlabs} and a conservative presumption of 0.5\% per surface for the AR coatings. Taking into account the lenses  angles of incidence we obtain the throughput of 70.6\% at 700 nm for the WFS branch. 

\subsection{Scientific payload branch}
\label{sec:science}

\begin{figure} [ht]
   \begin{center}
   \begin{tabular}{c} 
   \includegraphics[width=13cm]{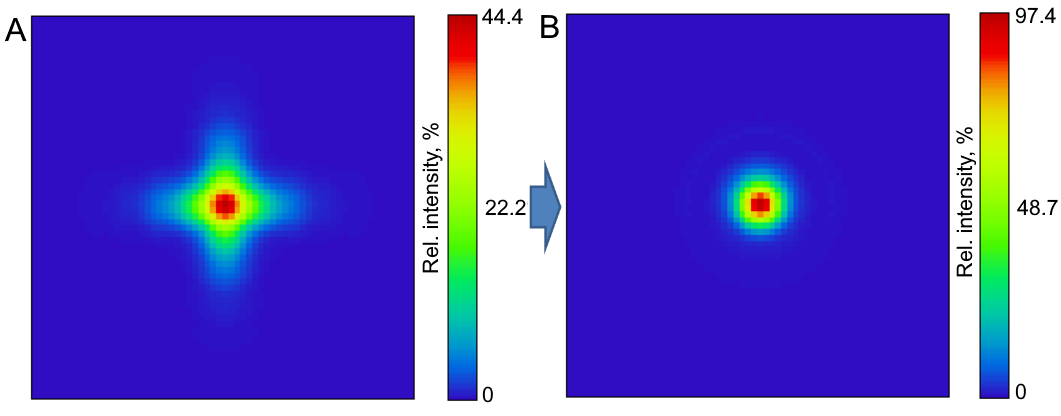}
   \end{tabular}
   \end{center}
   \caption[sciencePSF] 
   { \label{fig:sciencePSF} 
PSF in the science branch: A - initial PSF with the projection system residual aberrations, B -  PSF in the science branch with corrected static aberrations.}
   \end{figure}

After correction of the static aberrations we can analyze the expected performance in the science branch. If we apply the same DM profile to it , we can show the gain in image quality from Strehl ratio of 44.4\% (Fig.~\ref{fig:sciencePSF},A) to 97.4\% (Fig.~\ref{fig:sciencePSF},B) in the polychromatic PSF. This demonstrates that the non-common path aberrations between the two branches like the chromatism in the modulation mirror relay lenses are negligible. So with the simplest optical components we can get fairly close to the diffraction limit just by using a small fraction of the DM stroke. 

In addition we can provide a rough estimation of the science branch throughput. Under the same assumptions about the coatings we obtain 53.3\%, 81.5\% and 78.5\% at 400, 550 and 640 nm, respectively. The transmission curve of the commercial dichroic splitter has a sharp drop closer to the working range limit at 680 nm. In general, the throughput can be significantly improved with dielectric mirrors and/or customized dichroic.  

\subsection{Expected performance gain}
\label{sec:gain}

Finally, we provide an estimate for the expected performance of the AO bench with the atmospheric conditions typical for the SAO site \cite{Shikhovtsev2020} and the BTA parameters\cite{Kukushkin2016}:
\begin{itemize}
    \item Fried parameter for the atmospheric turbulence $r_0=8 cm$;
    \item Wind speed $V_0 =8 m/s$;
    \item Telescope primary mirror diameter  $D = 6 m$;
    \item Central obscuration $Obs=0.33$;
    \item Number of actuators in a single line \textit{17} or \textit{24};
    \item Loop frequency  $f=500 Hz$;
    \item Loop delay $t_d=2 ms$;
    \item Target - fiber core $100 \mu m$ in diameter sampled at least by 10 elements $10 \mu m$ each.
\end{itemize}

\begin{figure} [ht]
   \begin{center}
   \begin{tabular}{c} 
   \includegraphics[width=13cm]{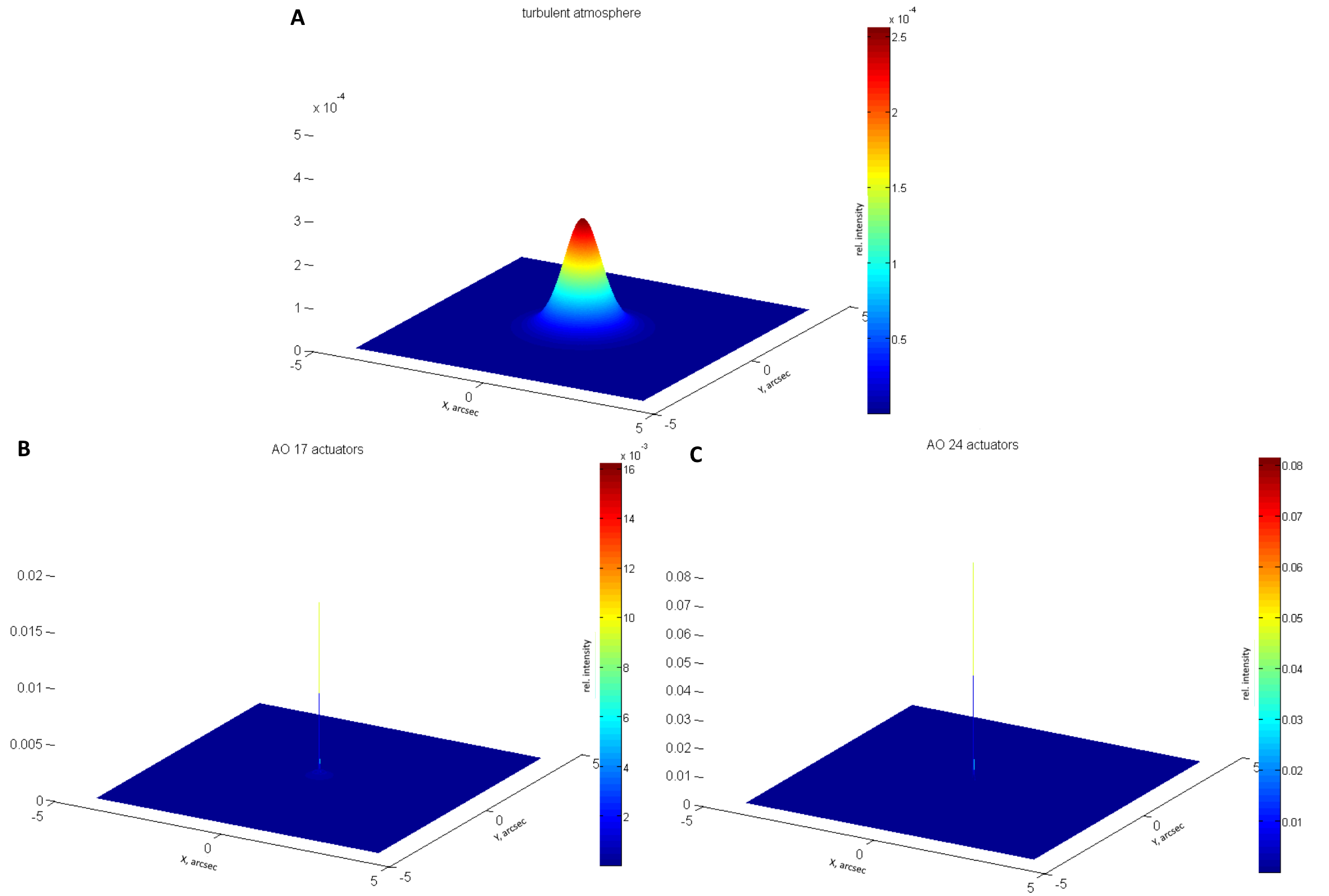}
   \end{tabular}
   \end{center}
   \caption[AOpsf] 
   { \label{fig:3dPSF} 
Computed PSF with the atmospheric turbulence: A - uncorrected, B - corrected by AO with 17x17 actuators DM, C - corrected by AO with 24x24 actuators DM.}
   \end{figure}

Using a general algorithm \cite{Fauvarque2016} and the corresponding modelling tools as well as the  approaches previously used by our colleagues \cite{Fetick2019, Beltramo2020} for modelling of point spread functions (PSFs) in adaptive-optics corrected systems, we obtain the expected PSFs of the designed bench. Fig.~\ref{fig:3dPSF} shows the intensity distribution normalized to the diffraction efficiency case for the uncorrected turbulence, low spatial sampling with 17x17 actuators and enhanced sampling with 24x24 actuators in the sub-plots A,B and C, respectively. Note that the DM's with 24 and 17 actuators have different diameters, but the pupil projection can fit within another DM after a moderate change of the concave mirrors positions.

One can see that by using the AO bench it becomes possible to compensate a significant fraction of the atmospheric turbulence and increase the Strehl ratio from 0.025\% to 1.6\% or 8\% at 540 nm depending on the number of actuators.

\begin{figure} [ht]
   \begin{center}
   \begin{tabular}{c} 
   \includegraphics[width=10cm]{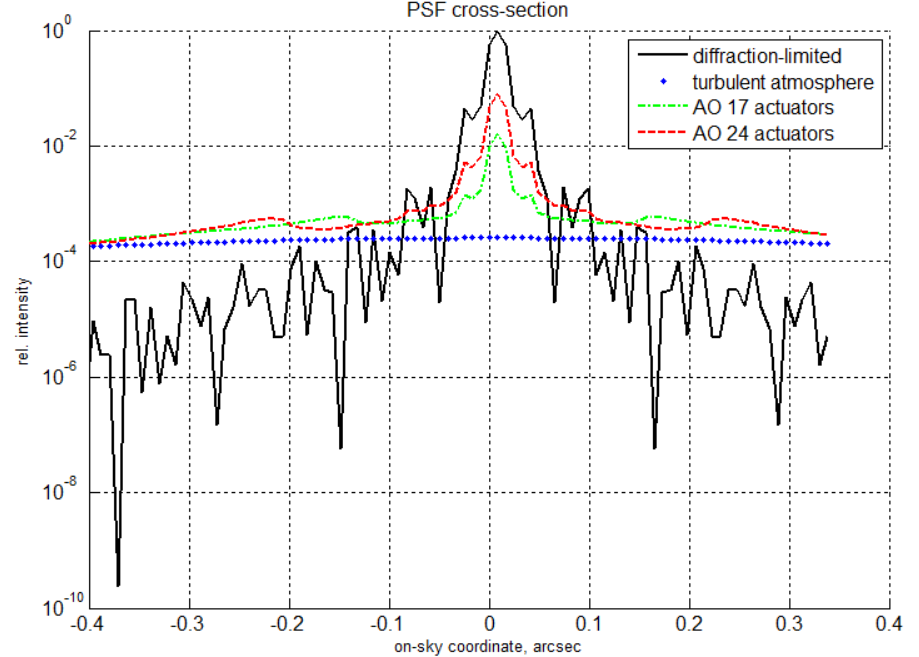}
   \end{tabular}
   \end{center}
   \caption[crossPSF] 
   { \label{fig:crossPSF} 
PSF cross-section at the spectrograph fiber input plane.}
   \end{figure}

To visualize the difference between the different modes better we plot the PSF cross-section in log-scale in Fig.~\ref{fig:crossPSF}. Though the image quality remains too low to use the Strehl ratio as the main metric. So, as the final estimate we use the relative energy concentration in the $100 \mu m$ diameter target corresponding to the spetrograph's fiber end. This simple analysis shows that use of the AO bench with 17x17 actuators DM allows to increase the energy concentration by a factor 8.37, while for the 24x24 actuators case this figure equals to 69.47. Then one can expect that the instrument sensitivity will be increased accordingly.

\section{CONCLUSIONS}
\label{sec:concl}  

Thus in the present paper we proposed an optical design of adaptive optics system for a high-resolution echelle spectrograph at 6-m telescope. It uses a pyramid wavefront sensor and is based on relatively simple optical components and available commercial active components. The entire design fits around the spectrograph's input unit and restores the nominal beam parameters. The flux losses due to introduction of the AO in the middle of the working waveband are estimated at the level of 18.5\%, but may be minimized by use of customized coatings. The static aberrations of the bench can be compensated by the deformable mirror with use of less than 2.7\% of its' nominal stroke.

Application of this AO system with the typical atmospheric conditions can allow to increase the energy concentration in the spectrograph input fiber core by a factor of 69.47 thus significantly increasing its' capacity to study faint objects.

\section*{ACKNOWLEDGMENTS}       
 
We would like to warmly thank our colleague Romain Fetick from LAM for his help with the simulations of atmospheric turbulence and its' correction.

GV thanks the grant of the Ministry of Science and Higher Education of the Russian Federation no. 075-15-2020-780 (no. 13.1902.21.0039).  

\bibliography{main} 
\bibliographystyle{spiebib} 

\end{document}